**Clever mothers balance time and effort in parental care – a study on free-ranging dogs**

Manabi Paul, Shubhra Sau, Anjan K. Nandi and Anindita Bhadra[*]

*Behaviour and Ecology Lab, Department of Biological Sciences, Indian Institute of Science Education and Research Kolkata, India.* (M. P., S. S., A. B.)

*Department of Physical Sciences,*
*Indian Institute of Science Education and Research Kolkata, India* (A. K. N.)

[*]Behaviour and Ecology Lab, Department of Biological Sciences,
Indian Institute of Science Education and Research Kolkata
Mohanpur Campus, Mohanpur,
PIN 741246, West Bengal, India.
tel. 91-33- 66340000-1223
fax +91-33- 25873020
e-mail: abhadra@iiserkol.ac.in

**Abstract:**

Mammalian offspring require parental care, at least in the form of suckling during their early development. While mothers need to invest considerable time and energy in ensuring the survival of their current offspring, they also need to optimize their investment in one batch of offspring in order to ensure future reproduction and hence lifetime reproductive success. Free-ranging dogs live in small social groups, mate promiscuously, and lack the cooperative breeding biology of other group living canids. They face high early life mortality, which in turn reduces fitness benefits of the mother from a batch of pups. We carried out a field based study on free-ranging dogs in India to understand the nature of parental care provided by mothers at different stages of pup development. Using behavioural patterns of mother-pup interactions, we draw up a timeline of pup ontogeny. Our analysis reveals that mothers cleverly reduce investment in energy intensive active care and increase passive care as the pups grow older, thereby keeping overall levels of parental care more or less constant over pup age.

[*]*Correspondent:* abhadra@iiserkol.ac.in

**Introduction:**

Parental care is an essential part of mammalian development where parents, especially the mothers, invest their time, energy and resources to provide care to their offspring, enhancing the offspring's chances of survival (Clutton-Brock 1991; Woodroffe and Vincent 1994). Maternal care can be defined as the amount of resources invested by the mother to rear her current offspring at the cost of her own survival and future reproduction. Life history theory predicts that the mother should invest optimal amount of resources so that the energy expenditure for her current offspring remains balanced against the effects on her chances of survival and future reproduction (Evans 1990; Stearns 1992; Roff 1993). In mammals, lactation seems to be the most energetically demanding component of maternal care that can affect the mother's growth, survival and reproduction (Martin 1984; Oftedal 1985; Stearns 1992). According to Parental investment theory (Trivers 1974), a mother should adopt a conservative strategy that ensures her own future reproduction and survival by decreasing the allocation of resources to her current offspring (Festa-Bianchet and Jorgenson 1998). This trade-off in energy allocation could be expressed through the changes in behaviours during offspring rearing (Dall and Boyd 2004). Kleiman and Malcolm (1981) sorted maternal care into direct and indirect care (Kleiman and Malcolm 1981). Direct maternal care comprises of comparatively more energy consuming behaviours that require direct contact between the mother and offspring (Malcolm 1985). Lifetime reproductive success (LRS) of the mother can be achieved by adjusting the intensity of direct and indirect care between parturition and weaning.

Biparental care has mostly been reported in group-living social species including humans (Kleiman and Malcolm 1981; Woodroffe and Vincent 1994). Care by adults other than the parents has also been reported in cooperatively breeding species and is probably universal in the family Canidae (Malcolm 1985). Cooperative breeding is well known in pack-living social canids like wolves (*Canis lupus*), coyotes (*Canis latrans*), Arctic foxes (*Vulpes lagopus*), African wild dogs (*Lycaon pictus*), etc., where the dominant male-female pair suppresses the subordinates' reproduction (Estes and Goddard 1967; Mech 1970; Ewer 1973; Fox et al. 1975; Jennions and Macdonald 1994). In such species, the subordinates provide care to the offspring of the dominant pair, without reproducing themselves. In communally breeding species, however, multiple females breed at the same time and share dens or territories, and care is provided communally (Hoogland et al. 1989; Creel and Creel 1991).

Domestic dog (*Canis familiaris*) ancestors are thought to have diverged from modern wolf (*Canis lupus*) lineages 27,000 years ago (Skoglund et al. 2015). Dogs are a fascinating family of carnivores displaying a diverse range of social organization, from solitary living in human homes as pets to living in social groups in free-ranging conditions (Serpell 1995; Sen Majumder et al. 2013). Unlike wolves, dogs do not have any reproductive hierarchy and mate promiscuously (Sen Majumder and Bhadra 2015). Maternal care has been reported to be the predominant care received by the pups (Pal 2005), though allo-parental care is also possible (Pal 2005; Paul et al. 2014a). Pet dogs are mostly supervised by their masters, hence parental care is not an indispensable part of pups' survival and development. Indeed, for pet dogs, the pups are separated from the mothers quite early in their development (Serpell 1995). In free-ranging dogs, however, parental care is essential in the early stages of development for the survival of the pups, and in spite of the presence of parental care, the mortality rates in early life are as high as 81% (Paul et al. 2016). Lactating females devote substantial time and energy to suckle their pups till weaning (Paul and Bhadra, under review), thus imposing a high metabolic demand on the mothers (Rogowitz 1996). Early weaning and a small litter size would be advantageous for the mother only if pup mortality rates were low, thereby ensuring her fitness benefits. We carried out an extensive behavioural study to investigate if free-ranging dog mothers optimize investment in parental care to maximize their lifetime reproductive success.

**Materials and Methods:**

In a field based study we collected behavioural data from 15 dog groups having 22 mothers and their pups (78 pups), over a period of 11 weeks each, from the $3^{rd}$-$17^{th}$ weeks of pup age. The study was conducted in different parts of West Bengal, India, between October 2010 and February 2015, during the primary denning seasons. Females become more cautious or aggressive just after giving birth and prevent others from entering/ approaching the den. Since the pups do not emerge from the dens before the third week of age and the mother spends most of her time inside or around the dens, it was impossible to carry out observations on them during this period. Each group was observed for two morning (0900-1200h) and two evening (1400-1700h) sessions spread over blocks of two weeks. Each three hour observation session consisted of 18 scans and 18 all occurrences sessions (AOS), amounting to a total of 8712 scans of one minute each and 8712 AOS of five minutes each for all the 22 mother-litter sets. Any form of mother-pup interactions that could increase the chances of pup survival were recorded as maternal care. Maternal care was sorted into active (that require direct contact with the pup, thus demanding more energy) and passive categories (not necessary to be in contact with the pup) (Supporting Information S1). Active maternal care that were brief in duration and/or less frequent in occurrence, such as suckling, allo-grooming, regurgitation, food offering and den cleaning were considered as the "events" and the other behaviours as "states" (Altmann 1974). The methods of the experiment were approved by the IISER Kolkata animal ethics committee, and met with the guidelines on animal ethics of the Government of India. All work reported here was purely observation based, and no animals were harmed or disturbed during the observations and followed ASM guidelines (Sikes et al. 2011)

Statistics:

For statistical analysis, we used StatistiXL 1.10 and Statistica version 12 and R statistics. We used "nlme" package in R statistics (Team R. 2015) for "linear mixed-effects model analysis". For the generalized linear mixed effect models we incorporated the group identity and the year of data collection as the "random effects". Predictor variables such as the pup age and current litter size (incorporating the change in litter size over time due to mortality) of the mother were added in the models as "fixed effects". As the residuals play an essential part in the model validation process, we did the "Bartlett test of homogeneity of variances" to check for homoscedasticity, separately for two predictor variables i.e. age and litter size (Zuur et al. 2009). The models where predictor variables were observed to violate the assumption of homogeneity, "varFixed" weight was added to fix the model (Zuur et al. 2009). We started with the full model, i.e., with all possible two-way interactions among the fixed effects. If the two-way interaction shown no significant effect on the response variable, we reduced the model using standard protocol of backward selection method and ended up with the optimal model.

Simple linear regression was used to check the relationship between a dependent variable and a single explanatory variable. Chi-square test was used to assess the goodness of fit between a set of observed and expected values.

**Results:**

Pup age and current litter size of the mother independently showed significant effect on the proportion of time spent by the mother in total maternal care (Linear mixed effect model: $P =$ age: 0.03, current litter size: 0.001) (Table 1a) (Supporting Information S2). However, the time spent in active maternal care was regulated by a combination of pup age and mother's current litter size (Linear mixed effect model: pup age*mother's current litter size: $P = 4e^{-04}$) (Table 1b) (Supporting Information S3). A 3d surface plot of litter size, pup age and active care revealed that the mothers having small litters showed the highest amount of active care at the early stages of pup development. However, for the same age of pups, care decreased with an increase in litter size, suggesting that the mothers regulated the investment in active care according to the litter sizes they had to nurture (Fig. 1a). Hence pups having fewer siblings tend to receive higher amount of active maternal care at their early stages of life (Linear mixed effect model: pup age*mother's current litter size: $P < 0.0001$) (Table 1c) (Supporting Information S4) (Fig. 1b).

Unlike active care, the proportion of time spent by the mother in passive care only depended on the pup age but not on the litter size (Linear mixed effect model: pup age: $P < 0.0001$, current litter size: $P = 0.85$) (Supporting Information S5). As the pups grew older, the mothers struck a balance between the proportion of time they spent in active and passive care towards their pups by decreasing the first and increasing the latter (Linear regression comparison: $F_1$= 207.767, $P <$ 0.0001) (Fig. 2).

The mother also distributed her time unequally among the various active care behaviours. With the pups growing older, the mothers reduced their investment in behaviours like suckling allo-grooming and piling up with pups (Linear regression: $R^2$= 0.805, *std. β*= -0.897, $P < 0.0001$); and invested more time in play and protective behaviours (Linear regression: $R^2$= 0.792, *std. β*= 0.890, $P < 0.0001$) (Fig. 3).

Suckling and allogrooming occupied (33 ± 16)% of the mothers' time during the third week of pup age, and reduced to zero percent by the 17th week of pup age. All the 22 mothers were observed to suckle and allogroom their own pups, with a change in the rate (frequency per hour) of these behaviours over pup age (Variation within mothers over pup age: ANOVA: suckling: $F$= 9.82, $P < 0.0001$; allo-grooming: $F$= 6.12, $P < 0.0001$, variation between mothers: ANOVA: suckling: $F$= 1.212, $P$= 0.25; allo-grooming: $F$= 1.66, $P$= 0.05, Fig. 4a, b). 10 of the mothers were observed to provide regurgitated food to their pups on being solicited. Regurgitation was recorded as early as the 5th week of pup age and continued till the 15th week (Fig. 4c). In 8 mother-litter groups, mothers were observed to offer scavenged food to her own pups from the 9th to 16th week of pup age (Fig. 4d). 15 Mothers were observed to eat the fecal matter of their pups in order to clean their dens till the 6th week of pup age (Fig. 4e). Behaviours like regurgitation, food offering and den cleaning, though scattered and rare in occurrence, were seen in (54.5 ± 16.4) % of the total observed mother-litter groups. Mothers showed aggressive behaviours such as angry barking, chasing, growling, biting, fighting to the intruder dogs and remained alert of their presence when she stayed with her pups. All the 22 mothers were observed to protect their own pups by being aggressive to the intruders, and such aggression reduced only in and around the weaning period (7$^{th}$ and 8$^{th}$ week of pup age) (Variation within mothers over pup age: ANOVA: $F= 1.97, P = 0.02$) (Fig. 4f). Mothers showed conflict toward their pups not by aggression but by refusing most of the pups' suckling attempts as the pups grew older (Linear regression: $t = 2.715, P = 0.007$) (Fig. 4g). Mother-pup play interactions were recorded throughout the observation period irrespective of pup age (Variation within groups over pup age: ANOVA: $F = 1.16, P = 0.31$) (Fig. 4h), and it was seen that pups initiated play more frequently than their mothers (Two sample t test: $t = -4.402, P < 0.0001$); 73% of the total observed mother-pup play behaviours being pup-initiated.

The mothers did not show preferential treatment towards any individual pups in suckling initiation (Chi square: $X^2_1= 100, P < 0.0001$, Fig. 5a) and allogrooming (Chi square: $X^2_1= 29.160, P < 0.0001$, Fig. 5b) during the entire duration of observations, and thus can be considered to be quite impartial. However the rate (frequency per hour) of care received (in terms of suckling and allogrooming) by individual pups was regulated by a combination of pup age and the current litter size (Linear mixed effect model: pup age*current litter size: $P = 0.00124$) (Table 2) (Supporting Information S6).

Timeline of pup development:

We used the duration of the different parental care behaviours to draw up a timeline of pup development. Suckling and allogrooming became less frequent from the 7$^{th}$ week of pup age, with most suckling attempts by pups being refused by the mother from this stage. Hence the 7$^{th}$ week of pup age marks the onset of weaning (Paul and Bhadra, under review). Mothers also stop den cleaning by eating their pups' fecal matter from the 7$^{th}$ week of pup age. However, mothers begin the process of weaning much earlier, i.e. the 5$^{th}$ week of pup age, by providing regurgitated food to the pups, thereby supplementing suckling and initiating the pups to scavenged food. All of these care giving behaviours stop at the 16$^{th}$ week of pup age, demarcating the time when the pups become independent of the mothers (Fig. 4). Hence we divide the developmental phase into two stages, pup (birth to 6$^{th}$ week of age) and juvenile (7$^{th}$-16$^{th}$ week of age). From the16$^{th}$ week of age until sexual maturity the individuals would thus be sub-adults.

**Discussion:**

Free-ranging dogs are known to have a facultatively social structure that is mostly influenced by their mating and pup rearing needs (Paul et al. 2014b, 2015; Sen Majumder and Bhadra 2015). The basic social unit during the denning season comprises of the mother and her pups. However other adults of the group may also invest directly or indirectly in the care of pups (Paul and Bhadra, in preparation, Paul et al. 2014a). In mammals, maternal care involves mother-offspring behavioural interactions that enhances the mother's reproductive success by ensuring the offspring's survival (Gubernick 1981). Disruption of maternal care often leads to adverse effects on the developing young (Gubernick 1981). In this study, maternal care was observed to be the most common form of care received by the pups of free-ranging dogs. Maternal care varied with pup age and litter size, with maximum care being provided by the mothers at the earliest stages of pup development. Though there was considerable variation in the care provided by individual mothers, there was a consistent trend in higher care being provided per pup when litter sizes were small. Thus pups with fewer siblings received more care in their early life.

Mothers invested a substantial amount of time in active care like suckling, allogrooming, piling up etc. at the very early stages of pup development, but adjusted their time activity budget to replace active care with passive care as the pups grew older. Suckling and allogrooming are both energy intensive behaviours (Rogowitz 1996; Carlini et al. 2004), and reducing investment in such behaviours can help the mother to replenish her energy reserves for future reproduction (Festa-Bianchet and Jorgenson 1998; Dall and Boyd 2004). A sharp decline in the rates of suckling and allogrooming were accompanied by offering of regurgitated and scavenged food to the pups in the pre-weaning period. Hence, while the mothers tried to wean the pups from milk, they also trained them on solid food, thus providing indirect care in the form of training. It is possible that pups learn to feed selectively on protein-rich food while scavenging through this process of training (Bhadra and Bhadra 2013). Pile sleep and piling up, i.e. grouping of mother and pups in a den, has also been observed to be an important maternal care behaviour that helps to maintain the body temperature of newborn pups (Lynch 1994). Such a close attachment between mother and pups was observed in the free-ranging dogs in the first couple of weeks of observation, and decreased as the pups grew older. Such close interactions are likely to have been even higher during the first two weeks after pup birth, when the mothers spent most of their time inside the dens. While these close contact behaviours reduced, play and protection increased with pup age, once again showing a switch from active to passive parental care.

We report three distinct phases in the ontogeny of pups, which is manifested through the changing maternal care patterns. Up to the 6$^{th}$ week of age, the pups receive ample active maternal care, and after this stage the onset of weaning occurs. The 7$^{th}$ – 13$^{th}$ week of age demarcates the "zone of conflict" between the mother and her pups (Paul and Bhadra, under review). The 7$^{th}$ week marks the end of the "pup" stage of life and the beginning of the "juvenile" phase. The juvenile phase continues into the post-weaning phase, when the mother-offspring interactions are more of social and physiological significance. By the 16$^{th}$ week of age, maternal care of all kinds ceases, and the juveniles enter the sub-adult phase of life, which is the phase leading up to sexual maturity and adulthood.

In mammals, age specific decline in female fertility followed by a sudden collapse in reproduction has been reported in various animals including dogs (Packer et al. 1998). Free-ranging dogs in India face high early life mortality of nearly 81%, leading to reduced fitness for the mothers. According to the predictions of life history theory (Evans 1990; Stearns 1992; Roff 1993), mothers could be expected to strike a balance between the cost and benefit of reproduction. Thus, the investment in one batch of offspring in terms of time and energy devoted to parental care (cost) should be adjusted to maximize their lifetime reproductive success (benefit) (Festa-Bianchet and Jorgenson 1998). Our study reveals that free-ranging dog mothers maximize their opportunity for future investment in reproduction by adjusting the active-passive care ratio over pup age, gradually shifting from more energy intensive care to less energy intensive care. This enables the mothers to provide prolonged care to their offspring at a relatively low investment. This might help to increase their fitness by providing protection to pups especially from interfering humans and by helping pups to learn efficient foraging strategies from their mothers through cultural transmission.

**Acknowledgements:**

The observations for this work were carried out by MP. MP, SS and AKN carried out the data analysis, AB supervised the work and co-wrote the paper with MP. The authors would like to thank Sreejani Sen Majumder, IISER Kolkata for her help during part of the fieldwork. This work was funded by projects from CSIR and SERB, India and supported by IISER Kolkata.

**References:**

Altman, J. 1974. Observational study of behavior: sampling methods. Behaviour 49:227–266.

Bhadra, A. and A. Bhadra. 2013. Preference for meat is not innate in dogs. Journal of Ethology 32:15–22.

Carlini, A., M. Marquez, H. Panarello and S. Ramdohr. 2004. Lactation costs in southern elephant seals at King George Island, South Shetland Islands. Polar Biology 27:266-276.

Clutton-Brock, T. 1991. The evolution of parental care. Princeton University Press, New Jersey.

Creel, S. and N. Creel. 1991. Energetics, reproductive suppression and obligate communal breeding in carnivores. Behavioral Ecology and Sociobiology 28:263-270.

Dall, S. and I. Boyd. 2004. Evolution of mammals: lactation helps mothers to cope with unreliable food supplies. Proceedings of the Royal Society of London, Series B 271:2049-2057.

Estes, R. D. and J. Goddard. 1967. Prey selection and hunting behavior of the African wild dog. Journal of Wildlife Management 31:52-70.

Ewer, R. 1973. The carnivores. Cornell University Press. USA.

Festa-Bianchet, M. and J. Jorgenson. 1998. Selfish mothers: reproductive expenditure and resource availability in bighorn ewes. Behavioral Ecology 9:144-150.

Fox, M., A. Beck and E. Blackman. 1975. Behavior and ecology of a small group of urban dogs (*Canis familiaris*). Applied Animal Ethology 1:119–137.

Gubernick, D. J. 1981. Parent and infant attachment in mammals. Pp. 243–305 in Parental care in mammals (D. J. Gubernick & P. H. Klopfer, eds.). Springer US.

Hoogland, J., R. Tamarin and C. Levy. 1989. Communal nursing in prairie dogs. Behavioral Ecology and Sociobiology 24:91–95.

Jennions, M. and D. Macdonald. 1994. Cooperative breeding in mammals. Trends in ecology & evolution 9:89–93.

Kleiman, D. and J. Malcolm. 1981. The evolution of male parental investment in mammals. Pp. 347–387 in Parental Care in Mammals (D. J. Gubernick & P. H. Klopfer, eds.). Springer US.

Lynch, C. 1994. Evolutionary inferences from genetic analyses of cold adaptation in laboratory and wild populations of the house mouse. Quantitative genetic studies of behavioral evolution 278-301.

Malcolm, J. R. 1985. Paternal care in canids. Integrative and Comparative Biology 25:853–856.

Martin, P. 1984. The meaning of weaning. Animal behaviour 32:1257–1259.

Mech, L. D. 1970. The wolf: ecology and behavior of an endangered species. Natural History Press. Garden City.

Oftedal, O. T. 1985. Pregnancy and lactation. Pp. 215–238 in Bioenergetics of Wild Herbivores (R.J.Hudson and R.G. White, ed.). CRC PressInc.,Boca Raton,FL.

Packer, C., M. Tatar and A. Collins. 1998. Reproductive cessation in female mammals. Nature 392:807-811.

Pal, S. 2005. Parental care in free-ranging dogs, Canis familiaris. Applied Animal Behaviour Science 90:31–47.

Paul, M., S. Sen Majumder and A. Bhadra. 2014a. Grandmotherly care: a case study in Indian free-ranging dogs. Journal of Ethology 32:75–82.

Paul, M., S. Sen Majumder and A. Bhadra. 2014b. Selfish mothers? An empirical test of parent-offspring conflict over extended parental care. Behavioural processes 103:17–22.

Paul, M., S. Sen Majumder, A. K. Nandi and A. Bhadra. 2015. Selfish mothers indeed! Resource-dependent conflict over extended parental care in free-ranging dogs. Royal Society Open Science 2:150580.

Paul, M., S. Sen Majumder, S. Sau, A. K. Nandi and A. Bhadra. 2016. High early life mortality in free- ranging dogs is largely influenced by humans. Scientific Reports 6:1–8.

Evans, R. M. 1990. The relationship between parental input and investment. Animal Behaviour 39:797–813.

Roff, D. 1993. Evolution of life histories: theory and analysis. Springer Science and Business media.

Rogowitz, G. L. 1996. Trade-offs in energy allocation during lactation. Integrative and Comparative Biology 36:197–204.

Sen Majumder, S., A. Bhadra, A. Ghosh, S. Mitra, D. Bhattacharjee, J. Chatterjee, A. K. Nandi and A. Bhadra. 2013. To be or not to be social: foraging associations of free-ranging dogs in an urban ecosystem. Acta Ethologica 17:1–8.

Sen Majumder, S. and A. Bhadra. 2015. When Love Is in the Air: Understanding Why Dogs Tend to Mate when It Rains. Plos One 10:e0143501.

Serpell, J. 1995. The Domestic Dog: Its Evolution, Behaviour and Interactions with People. Cambridge University Press, Cambridge.

Sikes, R. S., W. L. Gannon, and the Animal Care and Use Committee of the American Society of Mammalogists. 2011. Guidelines of the American Society of Mammalogist for the use of wild mammals in research. Journal of Mammalogy 92:235–253.

Skoglund, P., E. Ersmark, E. Palkopoulou and L. Dalen. 2015. Ancient wolf genome reveals an early divergence of domestic dog ancestors and admixture into high-latitude breeds. Current Biology 25:1515-1519.

Sterns, S. 1992. The evolution of life histories. Oxford University Press.

Team, R. 2015. R: A language and environment for statistical computing. R Foundation for Statistical Computing, Vienna, Austria.

Trivers, R. L. 1974. Parent-offspring conflict. American Zoologist 14:249–264.

Woodroffe, R. and A. Vincent. 1994. Mother's little helpers: Patterns of male care in mammals. Trends in ecology & evolution 9:294–7.

Zuur, A. F., E. N. Ieno, N. J. Walker, A. A. Saveliev and G. M. Smith. 2009. Mixed Effects Models and Extensions in Ecology with R. Springer Science+Business Media, LLC, 233 Spring Street, New York, USA.

**Tables**

a)

|  | Value | Std. Error | DF | t-value | P-value |
|---|---|---|---|---|---|
| (Intercept) | 0.9774950 | 0.11531726 | 169 | 8.476572 | 0.0000 |
| age | -0.0218620 | 0.01003326 | 169 | -2.178953 | **0.0307** |
| LS | -0.0987305 | 0.02951282 | 169 | -3.345342 | **0.0010** |

b)

|  | Value | Std. Error | DF | t-value | P-value |
|---|---|---|---|---|---|
| (Intercept) | 1.0513085 | 0.09673973 | 169 | 10.867392 | 0.0000 |
| age | -0.0799352 | 0.00898196 | 169 | -8.899528 | **0.0000** |
| LS | -0.1216618 | 0.02361754 | 169 | -5.151332 | **0.0000** |
| age*LS | 0.0092408 | 0.00257809 | 169 | 3.584355 | **4e-04** |

c)

|  | Value | Std. Error | DF | t-value | P-value |
|---|---|---|---|---|---|
| (Intercept) | 0.5574674 | 0.04028909 | 169 | 13.836682 | 0.0000 |
| age | -0.0377047 | 0.00383511 | 169 | -9.831449 | **0.0000** |
| LS | -0.0905445 | 0.01000378 | 169 | -9.051032 | **0.0000** |
| age*LS | 0.0052365 | 0.00110413 | 169 | 4.923760 | **0.0000** |

**Table 1:** Results of the linear mixed effect models for the effect of pup age (age) and mother's current litter size (LS) on the proportion of time spent by the mother in maternal care, shown toward her pups. (a) Age and current litter size independently affect the total care shown by the mothers. (b) Results showing significant effect of age and litter size and the interaction between age and litter size over the proportion of time spent by the mother in active care. (c) Active care received by individual pup depends on the combined effect of their age and current litter size.

**Table 2.**

| | Value | Std. Error | DF | t-value | P-value |
|---|---|---|---|---|---|
| (Intercept) | 9.61189 | 0.91536 | 169 | 10.501 | 2.22e-16 |
| age | -0.73337 | 0.08936 | 169 | -8.207 | **6.66e-14** |
| LS | -1.12415 | 0.23868 | 169 | -4.710 | **7.24e-06** |
| age*LS | 0.08342 | 0.02542 | 169 | 3.282 | **0.00124** |

**Table 2:** Results of the linear mixed effects model for the rate of care received by individual pup (in terms of suckling and allogrooming). The rate of care received is regulated by a combination of pup age and their current litter size.

**Figures**

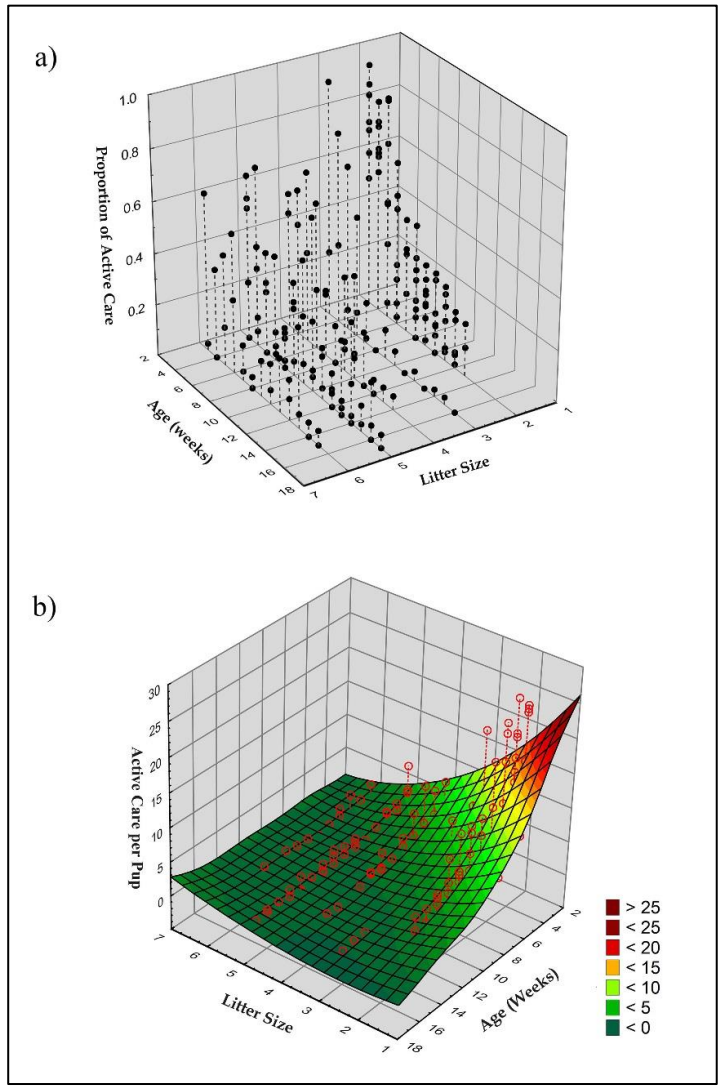

**Figure 1:** 3D graphical representation of the combined effect of pup age and current litter size on the active care shown by the mother. (a) 3D scatter plot showing that mothers having small litters show the highest amount of active care at the early stages of development of the pups. However the care decreases for the same pup age if the mother has a large litter. (b) 3D surface plot

showing that the pups having a fewer siblings tend to receive higher amount of active maternal care at the early stages of their life.

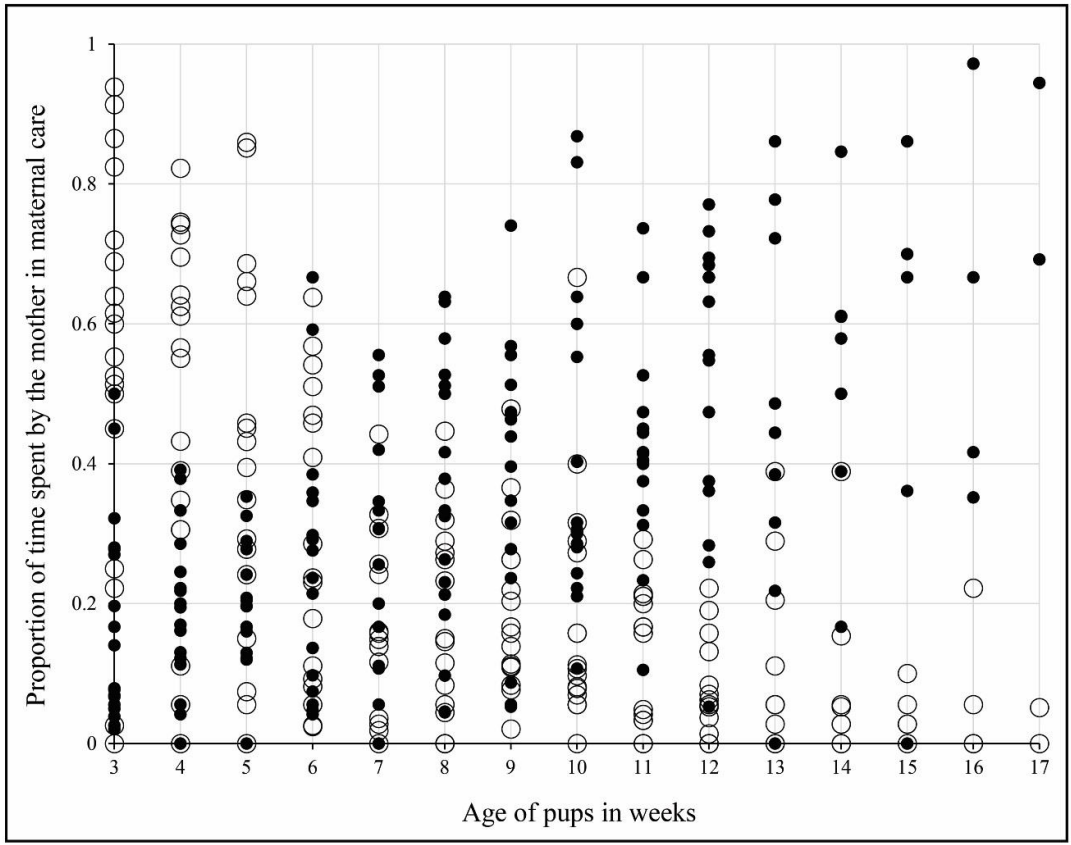

**Figure 2:** Scatter plot showing that as the pups grow older, the mothers strike a balance between the proportion of time spent in active and passive care towards their pups by decreasing the first and increasing the latter. Empty circles represent active care and solid circles represent passive care.

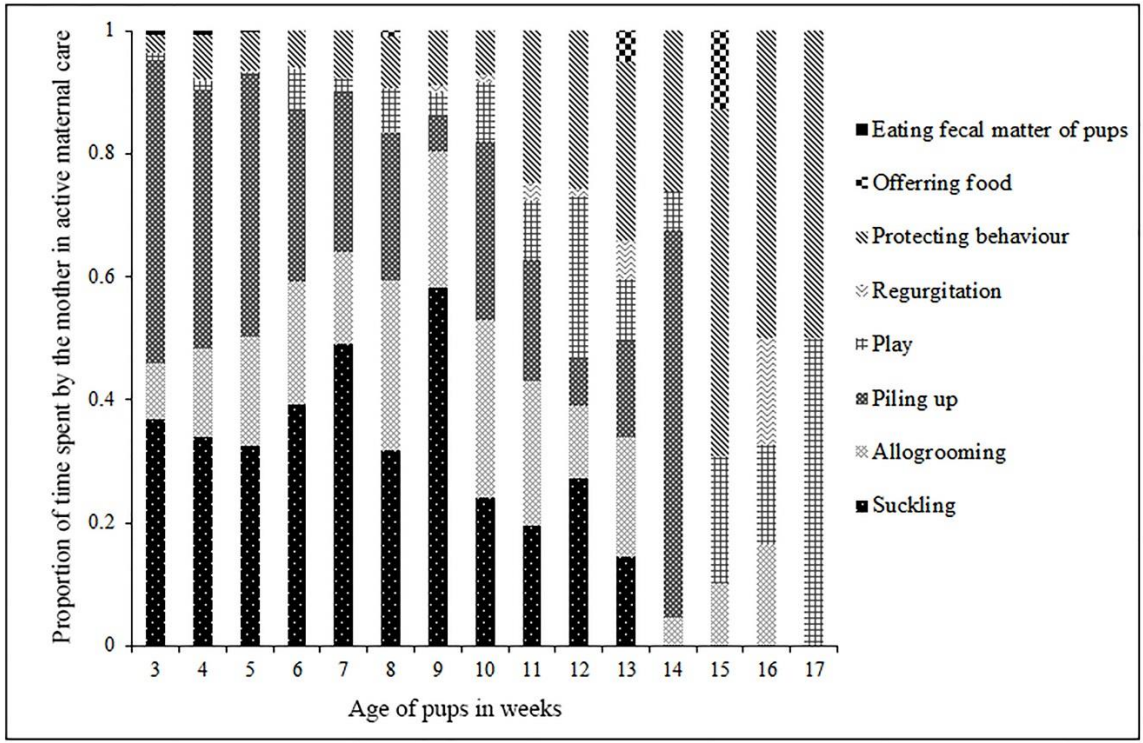

**Figure 3:** Stacked bar diagram showing that the mothers distribute their time unequally among the various active care behaviours. The mothers reduce their investment in behaviours like suckling, allogrooming and piling up with pups and invest more time in play and protective behaviours with increasing pup age.

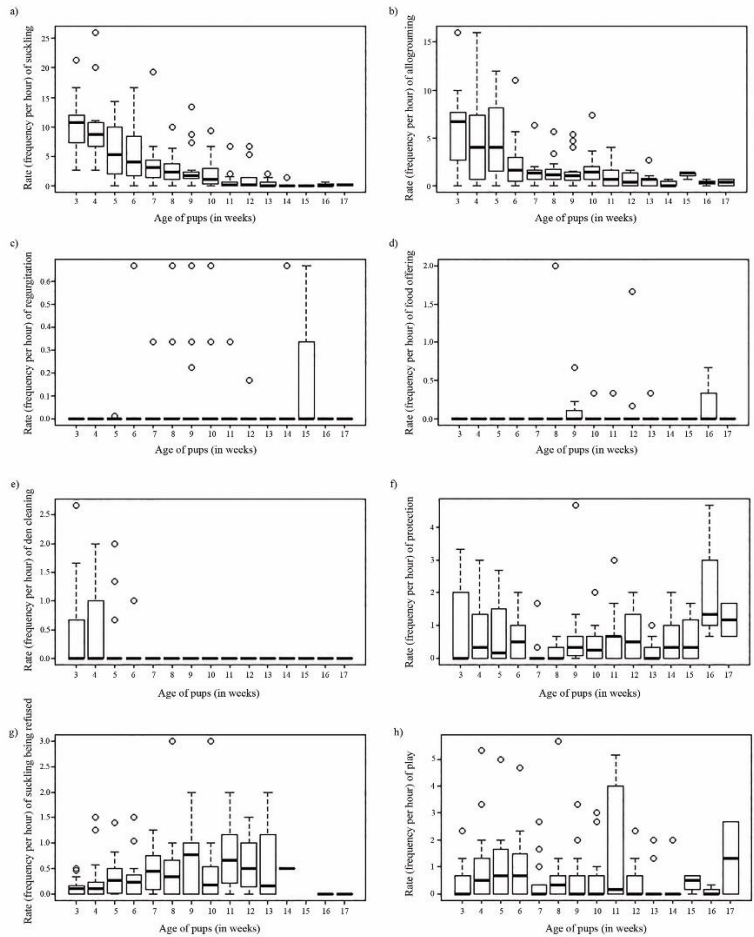

**Figure 4:** Box-whisker plots showing the rate (frequency per hour) of active care shown by the mothers toward their pups over increasing pup age. (a) Rate of suckling, (b) rate of allogrooming, (c) rate of regurgitation, (d) rate of food offering, (e) rate of den cleaning by eating the fecal matter of pups, (f) rate of protection, (g) rate of suckling being refused by the mother and (h) the rate of play interactions between mother and pups over increased pup ages.

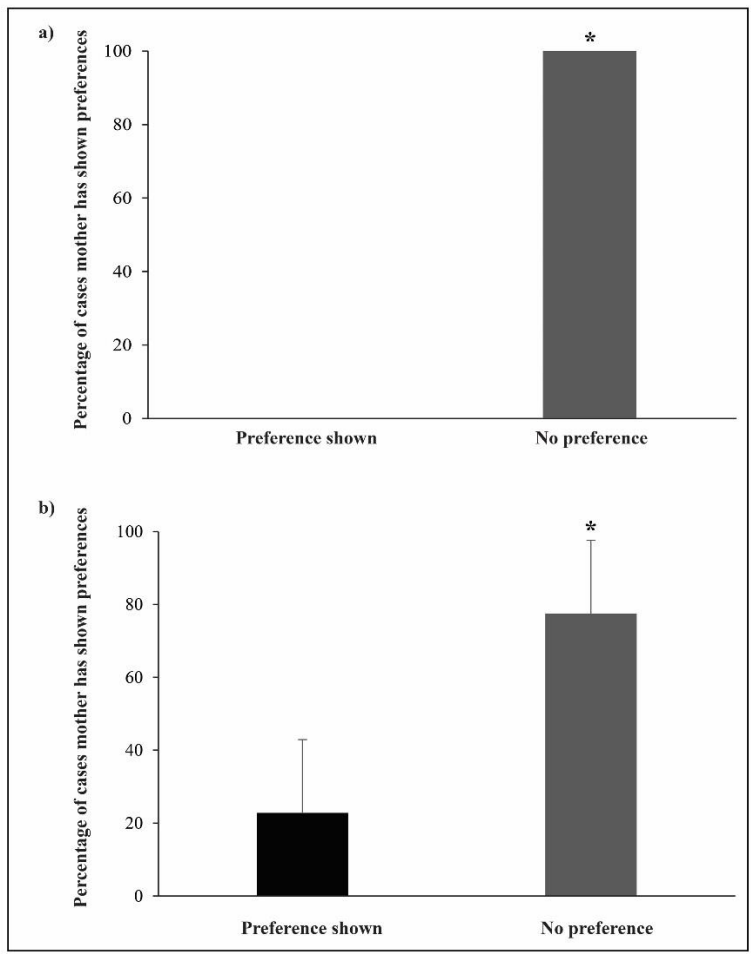

**Figure 5:** Bar diagram showing (a) the preference shown by the mother while she suckles her pups and (b) the preference shown by the mother while she allogrooms her pups.

**Supporting Information:**

**Supporting information S1**.—A list of active and passive maternal care behaviours observed. Each behaviour has a two letter code along with their descriptions.

| Mother-pup affiliative interactions (Maternal care) | | | | | |
|---|---|---|---|---|---|
| Active maternal care | | | Passive maternal care | | |
| Code | Behaviour | Description | Code | Behaviour | Description |
| AG | Allo groom | Groom own pups | DW | Drink water | Drink water |
| CK | Suckle | Feed milk to pups | ET | Eat food | Eat food |
| EF | Eat fecal matter | Mother eat fecal matter of her pups to clean her den | FS | Food search | Forage in search of food |
| OF | Offer | Offer food to pups | GR | Groom | |
| PL | Play | Play with pups | LG | Scratching body | Scratch her body by legs |
| PS | Pile sleep | Sleep together forming a pile | LI | Licking | Lick her body |
| PU | Pile Up | Sit together forming a pile | OT | Others | Mother rests yet aware of her pups |
| TH | Threat | Threat the intruders | | | |
| VM | Vomit | Vomit to provide regurgitated food to pups | | | |

**Supporting information S2.**—Details of the linear mixed effect model that shows the effect of pup age and their current litter size on the proportion of time spent in total care by the mother.

In order to check the effect of both the predictor variables i.e. pup age and mother's current litter size, we ran a "linear mixed effect model" incorporating the predictor variables as the "fixed effects" whereas the proportion of time spent by the mother in total care was included in the model as the "response variable". We collected the data on maternal care from 15 different dog groups that have 22 mother-litter units, over a span of 5 years (2010-2015). Hence the identity of each mother-litter units (fgr) and the year of data collections (fyr) were incorporated in the model as the "random effects". A Gaussian distribution was considered for the response variable in the model. We started with the full model (mod1), i.e., with all possible two-way interactions among the fixed effects.

**Variables used in the model:**

Response variable:

Proportion of time spent by the mother in total care- **totpcprop**

Fixed effects:

Age of pups in weeks- **age**

Current litter size- **LS**

Random effects:

Group identity- **fgr**

Year of observation- **fyr**

Model: **mod1<- lme (totpcprop~ age*LS, random = ~1|fgr/fyr)**

**Results**

|  | Value | Std. Error | DF | t-value | p-value |
|---|---|---|---|---|---|
| (Intercept) | 0.9774950 | 0.11531726 | 169 | 8.476572 | 0.0000 |
| age | -0.0218620 | 0.01003326 | 169 | -2.178953 | **0.0307** |
| LS | -0.0987305 | 0.02951282 | 169 | -3.345342 | **0.0010** |
| age*LS | 0.0037576 | 0.00284882 | 169 | 1.318992 | 0.1890 |

The two-way interaction shown no significant effect on the response variable and hence we reduced the model using standard protocol of backward selection method and ended up with the optimal model (mod2).

Model: **mod2<- lme (totpcprop~ age + LS, random = ~1|fgr/fyr)**

**Results**

| | Value | Std. Error | DF | t-value | p-value |
|---|---|---|---|---|---|
| (Intercept) | 0.9774950 | 0.11531726 | 169 | 8.476572 | 0.0000 |
| age | -0.0218620 | 0.01003326 | 169 | -2.178953 | **0.0307** |
| LS | -0.0987305 | 0.02951282 | 169 | -3.345342 | **0.0010** |

**Supporting information S3.—** Details of the linear mixed effect model that shows the effect of pup age and their current litter size on the proportion of time spent in active care by the mother.

The predictor variables i.e. the "age" and "LS" were incorporated in the generalized linear mixed effect model to check their effect on the response variable i.e. the proportion of time spent in active care (acareprop) for 22 mother-pups unit, collected over a span of 5 years. Group identity of each mother-pups unit (fgr) and the year of data collections (fyr) were added in the model as the "random effects". We started with the full model (mod3), i.e., with all possible two-way interactions among the fixed effects.

Model validation

Since the residuals have an essential role in the model validation process, we did the "Bartlett test of homogeneity" of variances to check the presence of homoscedasticity in the model, separately for two predictor variables i.e. pup age (age) and current litter size (LS). P-value for the predictor "age" exhibited violation in the homogeneity assumption and it was fixed by adding "varFixed" as the "weight" in the model.

**Variables used in the model:**

Response variable:

Proportion of time spent by the mother in active care- **acareprop**

Fixed effects:

Age of pups in weeks- **age**

Current litter size- **LS**

Random effects:

Group identity- **fgr**

Year of observation- **fyr**

Model: **mod3<- lme(acareprop ~ age * LS, random= ~1|fgr/fyr, weights= varFixed(~age))**

**Results**

|             | Value      | Std. Error | DF  | t-value    | p-value |
|-------------|------------|------------|-----|------------|---------|
| (Intercept) | 1.0513085  | 0.09673973 | 169 | 10.867392  | 0.0000  |
| age         | -0.0799352 | 0.00898196 | 169 | -8.899528  | 0.0000  |
| LS          | -0.1216618 | 0.02361754 | 169 | -5.151332  | 0.0000  |
| age*LS      | 0.0092408  | 0.00257809 | 169 | 3.584355   | **4e-04** |

Here the two-way interaction between the age and LS shown significant effect on the response variable and hence we kept this model.

**Supporting information S4.**— Details of the linear mixed effect model that shows the effect of pup age and their current litter size on the active care received per pup.

Total active care shown by the mother for a particular group was divided equally among the pups that were observed to present for the respective week of pup age. Active care received per pup was incorporated into the generalized linear mixed effect model as the response variable. We wanted to check the effect of predictor age and LS on the amount of active care received per pup. Group identity of each mother-pups unit (fgr) and the year of data collections (fyr) were added in the model as the "random effects". We started with the full model (mod4), i.e., with all possible two-way interactions among the fixed effects.

Model validation

Since the residuals have an essential role in the model validation process, we did the "Bartlett test of homogeneity" of variances to check the presence of homoscedasticity in the model, separately for two predictor variables i.e. pup age (age) and current litter size (LS). P-value for the predictor "age" exhibited violation in the homogeneity assumption and it was fixed by adding "varFixed" as the "weight" in the model.

**Variables used in the model:**

Response variable:

Active care received per pup- **acareperpup**

Fixed effects:

Age of pups in weeks- **age**

Current litter size- **LS**

Random effects:

Group identity- **fgr**

Year of observation- **fyr**

Model: **mod4<- lme(acareperpup ~ age * LS, random= ~1|fgr/fyr, weights= varFixed(~age))**

**Results**

|              | Value      | Std. Error  | DF  | t-value    | p-value |
|--------------|------------|-------------|-----|------------|---------|
| (Intercept)  | 0.5574674  | 0.04028909  | 169 | 13.836682  | 0.0000  |
| age          | -0.0377047 | 0.00383511  | 169 | -9.831449  | **0.0000** |
| LS           | -0.0905445 | 0.01000378  | 169 | -9.051032  | **0.0000** |
| age*LS       | 0.0052365  | 0.00110413  | 169 | 4.923760   | **0.0000** |

Here the two-way interaction between the age and LS shown significant effect on the response variable and hence we kept this model.

**Supporting information S5.**— Details of the linear mixed effect model that shows the effect of pup age and their current litter size on the proportion of time spent in passive care by the mother.

The predictor variables i.e. the "age" and "LS" were incorporated in the generalized linear mixed effect model to check their effect on the response variable i.e. the proportion of time spent in passive care (pcareprop) for 22 mother-pups unit, collected over a span of 5 years. Group identity of each mother-pups unit (fgr) and the year of data collections (fyr) were added in the model as the "random effects". We started with the full model (mod5), i.e., with all possible two-way interactions among the fixed effects.

**Variables used in the model:**

Response variable:

Proportion of time spent by the mother in passive care- **pcareprop**

Fixed effects:

Age of pups in weeks- **age**

Current litter size- **LS**

Random effects:

Group identity- **fgr**

Year of observation- **fyr**

Model: **mod5<- lme(pcareprop ~ age * LS, random= ~1|fgr/fyr, weights= varFixed(~age))**

**Results**

|  | Value | Std. Error | DF | t-value | p-value |
| --- | --- | --- | --- | --- | --- |
| (Intercept) | 0.05735617 | 0.08746429 | 169 | 0.655767 | 0.5129 |
| age | 0.04428560 | 0.00786539 | 169 | 5.630439 | 0.0000 |
| LS | 0.00443542 | 0.02273035 | 169 | 0.195132 | 0.8455 |
| age*LS | -0.00370192 | 0.00225732 | 169 | -1.639963 | 0.1029 |

The two-way interaction shown no significant effect on the response variable and hence we reduced the model using standard protocol of backward selection method and ended up with the optimal model (mod6).

Model: **mod6<- lme(pcareprop ~ age + LS, random= ~1|fgr/fyr, weights= varFixed(~age))**

**Results**

| | Value | Std. Error | DF | t-value | p-value |
|---|---|---|---|---|---|
| (Intercept) | 0.05735617 | 0.08746429 | 169 | 0.655767 | 0.5129 |
| age | 0.04428560 | 0.00786539 | 169 | 5.630439 | **0.0000** |
| LS | 0.00443542 | 0.02273035 | 169 | 0.195132 | 0.8455 |

**Supporting information S6.**— Details of the linear mixed effect model that shows the effect of pup age and their current litter size on rate of (frequency per hour) of care received (in terms of suckling and allogrooming) by individual pups.

The predictor variables i.e. the "age" and "LS" were incorporated in the generalized linear mixed effect model to check their effect on the response variable i.e. the rate of (frequency per hour) of care received (in terms of suckling and allogrooming) by individual pups (carercvd) for 22 mother-pups unit, collected over a span of 5 years. Group identity of each mother-pups unit (fgr) and the year of data collections (fyr) were added in the model as the "random effects". We started with the full model (mod7), i.e., with all possible two-way interactions among the fixed effects.

**Variables used in the model:**

Response variable:

The rate of (frequency per hour) of care received (in terms of suckling and allogrooming) by individual pups - **carercvd**

Fixed effects:

Age of pups in weeks- **age**

Current litter size- **LS**

Random effects:

Group identity- **fgr**

Year of observation- **fyr**

Model: **mod7<- lme(carercvd ~ age * LS, random= ~1|fgr/fyr)**

**Results**

|  | Value | Std. Error | DF | t-value | p-value |
|---|---|---|---|---|---|
| (Intercept) | 9.61189 | 0.91536 | 169 | 10.501 | 2.22e-16 |
| age | -0.73337 | 0.08936 | 169 | -8.207 | **6.66e-14** |
| LS | -1.12415 | 0.23868 | 169 | -4.710 | **7.24e-06** |
| age*LS | 0.08342 | 0.02542 | 169 | 3.282 | **0.00124** |